\documentclass[aps,twocolumn,prb,preprintnumbers,amsmath,amssymb]{revtex4-2}
\allowdisplaybreaks[4]
\newcommand{\lsim} 
 {\ \raise.35ex\hbox{$<$}\kern-0.75em\lower.5ex\hbox{$\sim$}\ }
\newcommand{\gsim}
 {\ \raise.35ex\hbox{$>$}\kern-0.75em\lower.5ex\hbox{$\sim$}\ }

\usepackage{graphicx}
\usepackage{dcolumn}
\usepackage{bm}
\usepackage{amsmath}
\usepackage{times}
\usepackage[usenames]{color}
\usepackage{natbib}
\usepackage{ulem}

\begin{document}

\title{Two routes to quantum anomalous Hall states in altermagnets}

\author{Makoto Naka$^{1}$, Shuntaro Sumita$^2$, Yukitoshi Motome$^3$, and Hitoshi Seo$^{4}$}
\affiliation{$^{1}$School of Science and Engineering, Tokyo Denki University, Ishizaka, Saitama 350-0394, Japan}
\affiliation{$^{2}$Department of Basic Science, The University of Tokyo, Meguro, Tokyo 153-8902, Japan}
\affiliation{$^{3}$Department of Applied Physics, The University of Tokyo, Bunkyo, Tokyo 113-8656, Japan}
\affiliation{$^{4}$RIKEN Center for Emergent Matter Science, Wako, Saitama 351-0198, Japan}

\date{\today}

\begin{abstract}
We theoretically propose two possible routes to realizing quantum anomalous Hall states in altermagnetic materials.
We consider a minimal square-lattice Hubbard model with antisymmetric spin-orbit coupling associated with 
an orthorhombic crystal structure, which supports a topologically trivial altermagnetic state.
By incorporating Rashba-type spin-orbit coupling and external perturbations, we demonstrate that this trivial state can be turned into topological altermagnetic phases in two distinct ways.
The first route is driven by a staggered potential that breaks the symmetry connecting crystallographically
equivalent sublattices, leading to a topological altermagnetic ground state characterized by a quantized Hall conductivity $\left| \sigma_{xy} \right|=e^2/h$ and a Chern number $C=1$.
The second route is realized by applying a magnetic field perpendicular to the two-dimensional plane. 
The resulting topological state appears as a metastable state in the magnetic hysteresis loop, exhibiting a quantized Hall conductivity $\left| \sigma_{xy} \right|=2e^2/h$ associated with a Chern number $C=2$. 
We show that these topological transitions are accompanied by characteristic gap closings at the Brillouin-zone boundary, with the number of gap-closing points determining the Chern number.
Ribbon-geometry calculations reveal chiral edge states consistent with the bulk topological invariants and demonstrate distinct spin polarizations between the $C=1$ and $C=2$ states. 
Our results establish experimentally accessible routes to quantized anomalous Hall responses in altermagnets.
\end{abstract}

\maketitle

\section{Introduction}
The quantization of the anomalous Hall effect (AHE)~\cite{Haldane2010}, namely the quantum AHE (QAHE), is characterized by a quantized Hall conductivity in the absence of external magnetic fields. 
It has been extensively discussed in magnetic systems with broken time-reversal (TR) symmetry, including ferromagnets~\cite{Qiao2010}, magnetically doped topological insulators~\cite{Yu2010, Chang2013}, noncoplanar magnets~\cite{Ohgushi2000}, and orbital ferromagnets in moir\'e heterostructures~\cite{Serlin2020}.
In these systems, finite Berry curvature originates from spontaneous magnetization or fictitious magnetic fields associated with scalar spin chirality~\cite{Nagaosa2010}. 
Recently, another class of TR-symmetry broken magnets with collinear antiferromagnetic spin structure~\cite{Ahn2019, Naka2019, Hayami2019, Smejkal2020}, known as altermagnets~\cite{Smejkal2022}, has emerged as a promising platform for realizing the QAHE.

In altermagnets, opposite-spin magnetic sublattices are connected not by translation or inversion symmetry, but by other crystal symmetry operations, such as rotations, mirrors, glides, or screws, owing to alternating local environments in the lattice structure. 
This situation differs fundamentally from that of conventional antiferromagnets, where local TR symmetry breaking is compensated over the entire crystal by translation or inversion symmetry connecting opposite-spin sublattices. 
Consequently, altermagnets exhibit ``macroscopic'' TR symmetry breaking, similar to ferromagnets, despite having vanishing net magnetization. 
This distinctive symmetry property leads to spin-split electronic and magnon bands even in the absence of relativistic spin-orbit coupling (SOC)~\cite{Noda2016, Okugawa2018, Ahn2019, Naka2019, Krempasky2024}. 
It further underlies a variety of unconventional cross-correlation phenomena, including spin current generation~\cite{Naka2019, Naka2021, GonzalezHernandez2021}, the AHE and magneto-optical effects~\cite{Solovyev1997, Smejkal2020, Naka2020, Samanta2020, Naka2022, Zhou2025}, piezomagnetic responses~\cite{Aoyama2024, Yershov2024, Naka2025}, and unconventional superconductivities~\cite{Sumita2023, Zhang2023, Chakraborty2024, Sumita2025}.

Recently, the topological aspects of altermagnets have attracted growing attention, and several theoretical studies have proposed the QAHE in altermagnetic systems~\cite{Guo2023, Ma2024, GonzalezHernandez2025a, Antonenko2025, GonzalezHernandez2025b}.
Notably, Li {\it et al.} have proposed the QAHE in a canted antiferromagnetic system even before the concept of altermagnetism was introduced~\cite{Li2019}. 
Many of these proposals, however, rely on theoretically idealized settings, including specially designed lattice geometries such as two-dimensional checkerboard and Lieb lattices, or so-called Zeeman-type SOC, which neglects explicit spin mixing. 
Moreover, first-principles demonstrations have thus far been mostly restricted to artificial two-dimensional compounds.
In this context, microscopic routes to realizing the QAHE in representative altermagnetic materials, such as perovskite- and rutile-type transition metal compounds and organic conductors, remain to be clarified.

\begin{figure}
\begin{center}
\includegraphics[width=1\columnwidth, clip]{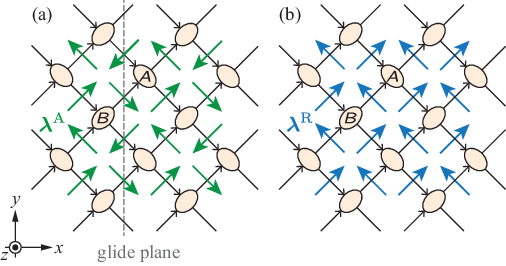}
\end{center}
\caption{
Schematic lattice structure and spatial patterns of the antisymmetric SOC vectors: (a) ${\bm \lambda}^{\rm A}_{ij}$ for an orthorhombic two-dimensional altermagnet and (b) Rashba-type SOC ${\bm \lambda}^{\rm R}_{ij}$. 
The green and blue solid arrows indicate the directions of ${\bm \lambda}^{\rm A}_{ij}$ and ${\bm \lambda}^{\rm R}_{ij}$, respectively, on the NN bonds associated with the electron hopping along the directions indicated by the black arrows. 
$A$ and $B$ denote the crystallographically equivalent sites in the unit cell, which are connected by the glide operation with respect to the $yz$ plane denoted by the dashed line.
}
\label{fig1}
\end{figure}

In this work, we investigate microscopic mechanisms for realizing the QAHE starting from a minimal model that satisfies the symmetry requirements for the AHE in altermagnets. 
By introducing experimentally accessible external perturbations, we demonstrate the quantization of the AHE and clarify its microscopic origins.

To this end, we introduce Rashba-type SOC to induce band inversion between the valence and conduction bands across the antiferromagnetic gap in insulating altermagnets. 
We then consider two kinds of perturbations: a staggered potential, which can be generated by uniaxial pressure, and a uniform external magnetic field. 
Both perturbations open gaps in the inverted band structure, leading to quantum anomalous Hall states.
We show that as a consequence, two distinct topological altermagnetic phases accompanied by quantized Hall conductivity with different Chern numbers are realized.
Interestingly, while one of these phases is stabilized as the ground state by the staggered potential, the other emerges as a metastable state during a magnetic-field-driven hysteresis process.

The remainder of this paper is organized as follows.
In Sec. II, we introduce a minimal Hubbard-type model for altermagnets, together with the external perturbations, and formulate the Hall conductivity and Chern number.
Section III presents the numerical results. 
In Sec. III A, we discuss the ground state in the absence of external fields. 
In Secs. III B and III C, we show the realization of the QAHE induced by the staggered potential and the magnetic field, respectively. 
In Sec. III D, we examine the corresponding edge states and characterize the topological nature of the obtained phases.
Finally, Sec. IV is devoted to discussions on the experimental relevance and generality of the proposed mechanisms.

\section{Model and Method}
We begin by introducing a minimal model for altermagnets.
The characteristic properties of altermagnets can be broadly classified into nonrelativistic effects, such as spin splitting and spin current generation, and relativistic effects, represented by the AHE. 
It has been pointed out that these phenomena originate from fundamentally different microscopic mechanisms and can therefore be modeled and treated separately~\cite{Naka2020, Naka2025rev, Solovyev2026, Solovyev2025}.
Since the QAHE studied in this work is a relativistic effect, it is important to distinguish its origin clearly from those of nonrelativistic effects. 
To this end, we adopt, as a starting point, a simplest model for altermagnets that can give rise to the altermagnetic AHE while avoiding unnecessary nonrelativistic spin splitting.

Satisfying these conditions, we consider a single-orbital Hubbard model on a square lattice with antisymmetric SOC that reflects the alternating local environments of the sublattices in altermagnets~\cite{Naka2020, Naka2025rev, Solovyev2026, Solovyev2025}, as illustrated in Fig.~\ref{fig1}(a):
\begin{align}
{\cal H}_{\rm ALM}
&= - t \sum_{ij,\sigma} c_{i\sigma}^\dagger c_{j\sigma}
+ U \sum_i n_{i\uparrow} n_{i\downarrow} \notag \\
&+ \sum_{ij,\sigma\sigma'}
\frac{i}{2}
({\bm \lambda}_{ij}^{\rm A}\cdot{\bm s})_{\sigma\sigma'}
c_{i\sigma}^\dagger c_{j\sigma'}.
\end{align}
Here $c_{i \sigma}$ ($c^{\dagger}_{i \sigma}$) and $n_{i\sigma}$ ($ = c^{\dagger}_{i \sigma}c_{i \sigma}$) are the annihilation (creation) and number operators of an electron with spin $\sigma$ at site $i$, respectively.
$t$ and $U$ represent the nearest-neighbor (NN) hopping integral and onsite Coulomb interaction, respectively. 
${\bm \lambda}^{\rm A}_{ij}$ denotes the antisymmetric SOC vector, which becomes nonzero on bonds lacking a local inversion center.
Although the configuration of ${\bm \lambda}^{\rm A}_{ij}$ is not unique for describing the crystal symmetries of altermagnets, 
we adopt the spatial arrangement shown in Fig.~\ref{fig1}(a), motivated by orthorhombic quasi-two-dimensional altermagnetic systems such as organic conductors~\cite{Naka2020} and layered or nonlayered perovskite-type compounds~\cite{Naka2022, Solovyev2026}.

In the presence of the ${\bm \lambda}^{\rm A}_{ij}$ term, the $A$ and $B$ sites are no longer connected by translation symmetry, and the unit cell consequently contains two sites related by a glide operation with respect to the $yz$ plane across the center of the NN bonds, as shown in Fig.~\ref{fig1}(a).
In addition, the mirror symmetries with respect to the $(110)$ and $(1\bar{1}0)$ planes, which are present in the simple square lattice, are lost due to ${\bm \lambda}^{\rm A}_{ij}$, resulting in orthorhombic crystal symmetry. 
In this way, the symmetry requirements for orthorhombic altermagnets are satisfied solely by ${\bm \lambda}^{\rm A}_{ij}$. 
In the absence of ${\bm \lambda}^{\rm A}_{ij}$, the present model reduces to the ordinary square-lattice Hubbard model, in which nonrelativistic spin splitting is absent even in the N\'eel-type antiferromagnetic state; to induce altermagnetic spin splitting, sublattice-dependent hopping terms would be required~\cite{Naka2019}, which are not considered here for the reason stated above.

Next, we take into account Rashba-type SOC, which is introduced in the same form as the ${\bm \lambda}^{\rm A}_{ij}$ term:
\begin{align}
{\cal H}_{\rm R} = \sum_{ij,\sigma\sigma'} \frac{i}{2} ({\bm \lambda}_{ij}^{\rm R}\cdot{\bm s})_{\sigma\sigma'} c_{i\sigma}^\dagger c_{j\sigma'},
\end{align}
where ${\bm \lambda}_{ij}^{\rm R}$ is another antisymmetric SOC vector with the spatial pattern shown in Fig.~\ref{fig1}(b). 
This term breaks inversion and mirror symmetries with respect to the $xy$ plane. 
More importantly, it induces band inversion across the antiferromagnetic gap, thereby providing the basis for the topological phase transitions discussed below.

\begin{figure*}
\begin{center}
\includegraphics[width=2.0\columnwidth, clip]{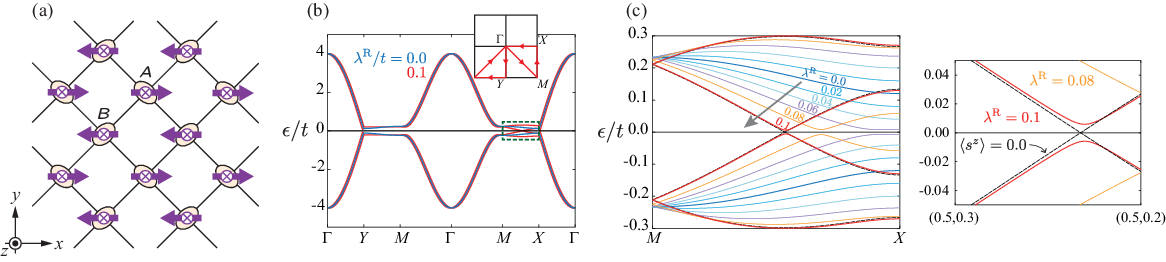}
\end{center}
\caption{(a) Schematic picture of an altermagnetic ground state of ${\cal H}_{\rm ALM}+{\cal H}_{\rm R}$ with the weak ferromagnetic moment along the $z$ axis.
(b) Energy band structures with and without the Rashba-type SOC term ${\cal H}_{\rm R}$ for $\lambda^{\rm R}/t = 0.0$ (blue) and $0.1$ (red), respectively.
The inset shows the path along the high-symmetry lines in the first BZ.
The dashed rectangle represents the region expanded in (c).
(c) Variation of the band structure near the insulating gap on the $X$-$M$ line with respect to $\lambda^{\rm R}/t$.
The dashed line denotes the band structure obtained from the self-consistent solution under the constraint of $\langle s^z \rangle = 0.0$ at $\lambda^{\rm R}/t=0.1$.
The right hand panel shows an enlarged view of the bands near the point node; $(k_x, k_y)$ denote the coefficients of the reciprocal vectors ${\bm b}_x$ and ${\bm b}_y$, respectively.
}
\label{fig2}
\end{figure*}

Finally, we introduce two kinds of external perturbations, namely a staggered potential and a uniform magnetic field, both of which can gap out the band crossings generated by the Rashba-induced band inversion. 
The staggered potential is given as
\begin{align}
{\cal H}_{\Delta}=\Delta\left(\sum_{i\in A}-\sum_{i\in B}\right)n_i,
\end{align}
where $\Delta$ is the amplitude of the staggered potential. 
In altermagnetic systems, such a staggered potential can naturally arise from uniaxial compression or tensile strain along the $\langle 110 \rangle$ direction, which deforms the local lattice environments around the $A$ and $B$ sites inequivalently~\cite{Naka2025}, as shown in Sec.~III B.

The uniform magnetic field is given by the Zeeman term for a field applied perpendicular to the two-dimensional plane as
\begin{align}
{\cal H}_{\rm Z} = - h \sum_{i} s_i^z,
\end{align}
where $h$ and $s^z_i = (n_{i\uparrow} - n_{i\downarrow})/2$ 
denote the magnetic-field strength and the $z$ component of the local spin operator, respectively.

We analyze the above Hamiltonian within the Hartree-Fock approximation, where the mean fields $\langle c^{\dagger}_{i\sigma} c_{i\sigma'} \rangle$ are determined self-consistently in the ground state. 
The $A$ and $B$ sites in the unit cell are treated independently. 

Using linear response theory, the intrinsic contribution to the Hall conductivity is calculated as 
\begin{align}
\sigma_{\mu \nu} = \frac{\hbar}{iN} \sum_{{\bm k}lm} &\frac{f(\epsilon_{{\bm k}l}) - f(\epsilon_{{\bm k}m})}{\epsilon_{{\bm k}l} - \epsilon_{{\bm k}m}} \notag \\
&\times \frac{[J^{\mu}(\bm k)]_{ml} [J^{\nu}(\bm k)]_{lm}}{\epsilon_{{\bm k}m} - \epsilon_{{\bm k}l} + i\gamma},
\label{sigma}
\end{align}
where $f(\epsilon_{{\bm k}l})$ is the Fermi distribution function for the Bloch eigenstate with wave vector $\bm k$ and band index $l$.
$[J^{\mu}(\bm k)]_{ml}$ is the matrix element of the $\mu$ component of the electric current operator between Bloch eigenstates, and $\gamma$ is the damping factor; $N$ is the total number of unit cells and the lattice constants are set to unity.

To characterize the topological properties of the obtained electronic states, we also calculate the Chern number~\cite{Thouless1982, Berry1985} defined by 
\begin{align}
C = \frac{1}{2\pi i} \int d\bm{k} F_z(\bm{k}), 
\end{align}
with
\begin{align}
F_z(\bm{k}) = \sum_{n \in \text{occupied}} [ \nabla_{\bm{k}} \times \langle u_n(\bm{k}) | \nabla_{\bm{k}} | u_n(\bm{k}) \rangle ]_z.
\end{align}
Here, $|u_n(\bm{k})\rangle$ is the periodic part of the Bloch wave function for the $n$th occupied band, and $F_z({\bm k})$ represents the Berry curvature summed over all occupied bands.
Since band degeneracies can occur among the occupied bands in the present system, we evaluate the Chern number using the non-Abelian extension of the Fukui-Hatsugai-Suzuki method on a discretized Brillouin zone~\cite{Fukui2005}.
The Chern number associated with the occupied band manifold is given by 
\begin{align}
C = \frac{1}{2 \pi i} \sum_{\bm k} \tilde{F}_{z}({\bm k}),
\end{align}
where, the lattice field $\tilde{F}_{z}({\bm k})$ is expressed as
\begin{align}
\tilde{F}_{z}({\bm k}) = \ln \frac{U_x({\bm k}) U_y({\bm k}+\delta{\bm k}_x)}{U_x({\bm k}+\delta{\bm k}_y) U_y({\bm k})}, 
\end{align}
with the link variable
\begin{align}
U_\mu({\bm k}) = \frac{1}{{\cal N}_\mu(\bm k)} \det \psi^\dagger({\bm k}) \psi({\bm k}+\delta {\bm k}_\mu), 
\end{align}
and its normalization factor
\begin{align}
{\cal N}_\mu(\bm k) = \left|  \det \psi^\dagger({\bm k}) \psi({\bm k}+\delta {\bm k}_\mu) \right|.
\end{align}
Here, $\delta {\bm k}_\mu$ is the vector connecting neighboring discretized $\bm k$ points along the $\mu$ direction. 
The matrix $\psi(\bm k)$ has dimensions $N_d \times N_b$ and composed of the column eigenvectors of the $N_b$ bands belonging to the occupied band manifold at the $\bm k$ point, with the dimension of the mean-field Hamiltonian $N_d$; for the present model, $N_d=4$ and $N_b=2$.

In the following, we demonstrate the two routes to realizing the QAHE in the present altermagnetic system. 
We take the hopping integral $t$ as an energy unit and, for simplicity, assume that the antisymmetric SOC vectors satisfy $|\lambda_x^{\rm A}| = |\lambda_y^{\rm A}| \equiv \lambda^{\rm A}$ and $|\lambda_x^{\rm R}| = |\lambda_y^{\rm R}| \equiv \lambda^{\rm R}$; $\lambda^{\rm A}$ and $\lambda^{\rm R}$ are treated as tunable parameters.
We focus on the case of half filling, corresponding to one electron per site, where an antiferromagnetic insulating state is stabilized for sufficiently large $U$ in the absecnce of SOC.

\section{Results}
\begin{figure*}
\begin{center}
\includegraphics[width=2.0\columnwidth, clip]{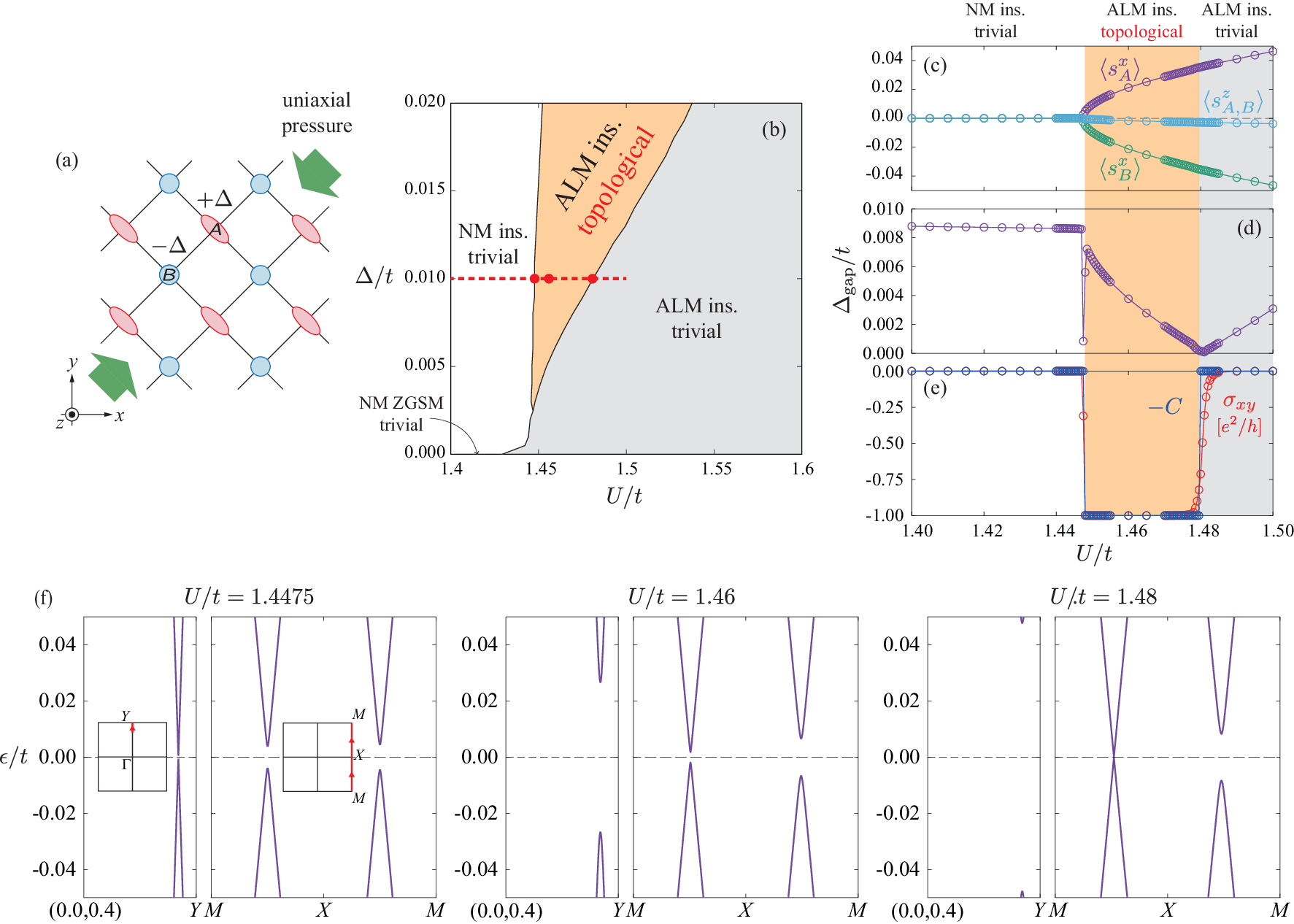}
\end{center}
\caption{
(a) Schematic illustration of the system with a staggered potential induced by uniaxial pressure.
(b) Ground-state phase diagram of ${\cal H}_{\rm ALM} + {\cal H}_{\rm R} + {\cal H}_{\Delta}$ in the $U$-$\Delta$ plane. 
NM, ALM, and ZGSM denote nonmagnetic, altermagnetic, and zerogap semimetal states, respeectively, and the orange region indicates the topological altermagnetic phase with the Chern number $\left| C \right| = 1$.
The dashed line represents the parameter path along which (c) the spin moments $\langle s^x_{A, B} \rangle$ and $\langle s^z_{A, B} \rangle$, (d) the energy gap between the valence and conduction bands, $\Delta_{\rm gap}$, and (e) the Hall conductivity $\sigma_{xy}$ and the Chern number of the occupied band manifold, $-C$, are calculated.
(f) Energy band structures near the gap for $\Delta/t = 0.01$ at $U/t = 1.4475$, $1.46$, and $1.48$, corresponding to the lower topological phase boundary, inside the topological altermagnetic phase, and the upper topological phase boundary, denoted by the circles in (b), respectively. 
The left and right subpanels show the band dispersions along the $\Gamma$-$Y$ and $M$-$X$-$M$ lines, respectively, demonstrating that the lower and upper topological transitions are driven by gap closings at different momenta.
$(k_x, k_y)$ denote the coefficients of the reciprocal vectors ${\bm b}_x$ and ${\bm b}_y$, respectively.
}
\label{fig3}
\end{figure*}
\subsection{Ground state without external fields}
We first discuss the ground state of the system in the absence of the staggered potential ${\cal H}_\Delta$ and the external magnetic field ${\cal H}_{\rm Z}$. 
For small $U$, the ground state of ${\cal H}_{\rm ALM}$ is a nonmagnetic zero-gap semimetal with nodes at the $X$ and $Y$ points in the first Brillouin zone. 
As $U$ exceeds a critical value, the system undergoes a phase transition to an altermagnetic state accompanied by a weak ferromagnetic moment, as schematically shown in Fig.~\ref{fig2}(a). 
In this phase, the antiferromagnetic moment is aligned along the $x$ axis, while a weak ferromagnetic component appears along the $z$ direction, with a vanishing $y$ component.

The obtained spin pattern is identical to those previously reported for the Hubbard model of an organic altermagnet $\kappa$-(BEDT-TTF)$_2X$~\cite{Naka2020} and for a recently proposed simplified model~\cite{Solovyev2025}, where a nonrelativistic spin splitting is absent, as mentioned above.
The spin pattern remains qualitatively unchanged upon introducing Rashba-type SOC term ${\cal H}_{\rm R}$.

Figure~\ref{fig2}(b) shows the change in the energy band structures induced by Rashba-type SOC for $U/t=1.5$ and $\lambda^{\rm A}/t = 0.1$. 
For $\lambda^{\rm R}/t = 0.0$, the energy gap is fully opened along the Brillouin-zone (BZ) boundaries, such as the $X$–$M$ and $Y$–$M$ lines, resulting in an insulating altermagnetic state. 
Upon introducing Rashba-type SOC with $\lambda^{\rm R}/t = 0.1$, the overall band structure and the bandwidth remain nearly unchanged, whereas the structures near the antiferromagnetic gap along the $X$-$M$ line exhibit a pronounced modification.

Figure~\ref{fig2}(c) shows the evolution of the local band structures near the gap along the $X$-$M$ line as $\lambda^{\rm R}/t$ is increased from $0.0$ to $0.1$. 
The Rashba-type SOC splits the degenerate bands, rapidly reduces the antiferromagnetic gap, and eventually induces band inversion between the valence and conduction bands. 
Consequently, a band-crossing point with a tiny gap appears along the $X$–$M$ line, as shown in Fig.~\ref{fig2}(c). 
This gap originates not from the antiferromangetic moment along the $x$ axis, but from the weak ferromagnetic moment along the $z$ axis that develops in the self-consistent solution.
Indeed, the black dashed line in Fig.~\ref{fig2}(c), which represents the band structure obtained under the constraint of $\langle s^z \rangle=0$ at $\lambda^{\rm R}/t=0.1$, exhibits a gapless point node. 
It should be noted that this tiny-gapped state is still topologically trivial with the Chern number $C = 0$, as discussed in the following.

\subsection{QAHE under a staggered potential}
We now investigate the case where a staggered potential is introduced into the topologically trivial altermagnetic state  discussed above; the Hamiltonian is given by ${\cal H}_{\rm ALM} + {\cal H}_{\rm R} + {\cal H}_{\Delta}$. 
Figure~\ref{fig3}(a) shows the ground-state phase diagram as a function of the Coulomb interaction $U$ and the staggered potential $\Delta$, with fixing $\lambda^A/t = \lambda^R/t = 0.1$. 
At $\Delta/t=0$ and small $U$, the Rashba SOC shifts the nodes away from the $X$ and $Y$ points, while the ground state remains a nonmagnetic zero-gap semimetal.
Upon increasing $U$, the system undergoes a second-order phase transition into an altermagnetic insulating phase, which is topologically trivial.

In the presence of a finite $\Delta/t$, the nonmagnetic semimetal is immediately gapped out, resulting in a topologically trivial nonmagnetic insulator.
For $\Delta/t \gtrsim 0.003$, a topological altermagnetic insulating phase emerges between the nonmagnetic insulating phase and the topologically trivial altermagnetic insulating phase. 
The stability region of this topological phase expands with increasing $\Delta$.
Figures~\ref{fig3}(c), \ref{fig3}(d), and \ref{fig3}(e) show the $U$ dependences of the local spin moments $\langle s^x_{A, B} \rangle$ and $\langle s^z_{A, B} \rangle$, the energy gap between the valence and conduction bands, $\Delta_{\rm gap}$, and the Hall conductivity $\sigma_{xy}$ and the Chern number of the occupied bands, $-C$, respectively, at $\Delta/t = 0.01$.
In the altermagnetic phases, the antiferromagnetic moment is aligned along the $x$ axis accompnanied by a weak ferromagnetic moment along the $z$ axis, as in the case without the staggered potential. 
All the moments increase smoothly with $U$ and exhibit no anomaly at the topological transition around $U/t=1.48$. 

In contrast, the energy gap $\Delta_{\rm gap}$ shown in Fig.~\ref{fig3}(d) exhibits clear signatures of the topological transitions.
Starting from the trivial nonmagnetic insulating phase, the gap sharply decreases and then reopens across the phase boundary to the topological altermagnetic phase at $U/t \simeq 1.45$. 
Within the topological altermagnetic phase, $\Delta_{\rm gap}$ decreases gradually with increasing $U$ and eventually goes to zero at $U/t \simeq 1.48$. 
Upon entering the trivial altermagnetic phase, the gap reopens and increases continuously.

Consistent with the gap-closing transitions described above, the Hall conducvitity $\sigma_{xy}$ shown in Fig.~\ref{fig3}(d) becomes quantized to $-e^2/h$ in the topological altermagnetic phase, whereas it remains zero in the two topologically trivial insulating phases.
Correspondingly, the Chern number of the occupied valence bands changes from $C=0$ to $C=1$ and then back to $C=0$ across the two gap-closing transiton boundaries.

Figure~\ref{fig3}(f) shows the band structures along the $\Gamma$-$Y$ and $M$-$X$-$M$ lines at the lower phase boundary ($U/t=1.4475$), inside the topological altermagnetic phase ($U/t=1.46$), and at the upper phase boundary ($U/t=1.48$). 
At the lower phase boundary, the energy gap closes along the $\Gamma$-$Y$ line, whereas it remains finite along the $M$-$X$ line. 
In contrast, at the upper phase boundary, the gap closes along the $M$-$X$ line in the $k_y < 0$ region, while it remains finite along the $\Gamma$-$Y$ line; the gap along the $X$-$M$ line in the $k_y>0$ region also remains open throughout the topological transition.
Thus, between these two transitions, finite gaps are present at both momenta, resulting in a topological altermagnetic insulating state with $C=1$.

\begin{figure}
\begin{center}
\includegraphics[width=1.0\columnwidth, clip]{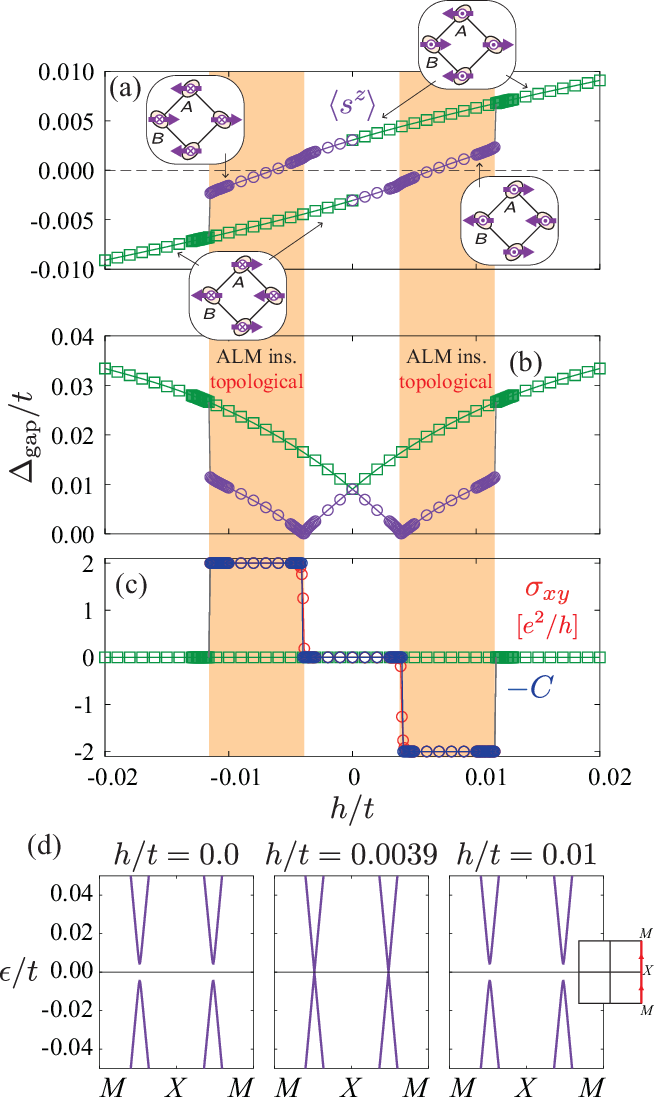}
\end{center}
\caption{
Magnetic-field dependences of  (a) the spin moment $\langle s^z_{A,B} \rangle$, (b) the energy gap $\Delta_{\rm gap}$, and (c) the Hall conductivity $\sigma_{xy}$ together with the Chern number of the occupied bands, $-C$, at $U/t=1.48$.
The insets in (a) show schematic spin configurations in the corresponding states.
The square and circle symbols represent the ground-state and metastable branches, respectively.
The shaded regions indicate the topological altermagnetic phases realized in the metastable branch.
(d) Energy band structures near the energy gap along the $M$-$X$-$M$ path at $h/t = 0$, $0.0039$, and $0.01$, corresponding to the trivial altermagnetic phase, the topological transition point, and the topological altermagnetic phase, respectively.
The inset illustrates the momentum-space path used for the calculation.
}
\label{fig4}
\end{figure}

\subsection{QAHE under a magnetic field}
Next, we turn to the effect of a uniform magnetic field applied perpendicular to the two-dimensional plane; the Hamiltonian is given by ${\cal H}_{\rm ALM} + {\cal H}_{\rm R} + {\cal H}_{\rm Z}$. 
Figure~\ref{fig4}(a) shows the field dependence of the uniform spin moment $\langle s^z \rangle \equiv \langle s^z_{A} \rangle=\langle s^z_{B} \rangle$ at $U=1.48$ and $\lambda^A/t = \lambda^R/t = 0.1$, along with the schematic spin patterns during the magnetization process.
The square symbols represent the ground-state solutions. 
For positive magnetic fields, the antiferromagnetic configuration with the $A$- and $B$-sublattice moments oriented downward and upward along the $x$ axis, respectively, is stabilized with positive $\langle s^z \rangle$. 
For negative magnetic fields, the opposite antiferromagnetic configuration is stabilized with negative $\langle s^z \rangle$.

Here, we focus on the metastable branch obtained when the magnetic field is applied opposite to the weak ferromagnetic moment. 
The circle symbols in Fig.~\ref{fig4}(a) show the evolution of $\langle s^z \rangle$ in the metastable altermagnetic state with negative (positive) $\langle s^z \rangle$ under weak positive (negative) magnetic fields. 
On the branch with the negative $\langle s^z \rangle$ at $h/t=0.0$, the magnitude of $\langle s^z \rangle$ gradually decreases as the field is increased in the positive direction, but remains finite at $h/t=0$, reflecting the weak ferromagnetism, as illustrated in the insets. 
Upon increasing the field to the positive side, $\langle s^z \rangle$ eventually vanishes near $h/t=0.005$, and then changes sign and grows parallel to the field direction. 
When the magnetic field exceeds a critical value of approximately $h/t \simeq 0.012$, the metastable solution ceases to exist and merges into the ground-state branch represented by the square symbols.
The closed curve consituted by these ground-state and metastable branches corresponds to the hysteresis loop of the uniform magnetization along the $z$ axis.

Figures~\ref{fig4}(b) and \ref{fig4}(c) show the magnetic-field dependences of the energy gap $\Delta_{\rm gap}$ and the Hall conductivity $\sigma_{xy}$ together with the Chern number $C$, respectively, for both the ground-state and metastable branches.
We first discuss the ground-state branch represented by the square symbols. 
As shown in Fig.~\ref{fig4}(b), the energy gap increases monotonically with increasing the magnetic field strengh, in both negative and positive directions. 
Therefore, no gap closing takes place and the system remains topologically trivial throughout the entire field range. 
Consistently, the Hall conductivity and the Chern number shown in Fig.~\ref{fig4}(c) both remain zero.

In contrast, the metastable branch exhibits a qualitatively different behavior. 
As shown in Fig.~\ref{fig4}(b), on the negative $\langle s^z \rangle$ branch at $h/t=0.0$ discussed above, the energy gap of the metastable solution, represented by the circle symbols, decrease continuouly from the the ground-state value and closes at $h/t \simeq 0.004$. 
Upon further increasing the magnetic field, the gap reopens and subsequently increases, before showing a discontinuous jump at $h/t \simeq 0.012$, where the metastable solution ceases to exist.
Correspondingly, the Hall conductivity and the Chern number shown in Fig.~\ref{fig4}(c) indicate that the system undergoes a topological transition at the gap-closing point. 
In the field range between the gap-closing point and the disappearance of the metastable state, the Hall conductivity is quantized to $\pm 2e^2/h$, accompanied by the Chern number $C=\mp 2$. 
These results demonstrate that topological altermagnetic phases are realized as metastable states during the magnetic hysteresis process.

Figure~\ref{fig4}(d) shows the evolution of the energy band structure for the metastable solution at $h/t>0$ near the gap along the $M$-$X$-$M$ line. 
At $h=0$, finite gaps are present on both sides of the $X$ point. 
With increasing the magnetic field, the gaps decrease and eventually closes at $h/t \simeq 0.004$. 
Upon further increasing the field, the gap reopens and the system enters the topological altermagnetic phase.
Unlike the staggered-potential-induced transition shown in Fig.~\ref{fig3}(e), where the gap closes only on one side of the $X$ point, the present transition is characterized by simultaneous gap closings on both sides of the $X$ point.

The difference originates from whether the twofold rotational symmetry about the midpoint of the nearest-neighbor bond is preserved. 
The staggered potential breaks this symmetry and makes the two magnetic sublattices inequivalent, whereas a perpendicular magnetic field preserves it. 
Consequently, the gap closes only on one side of the $X$ point in the staggered potential case, whereas symmetry requires simultaneous gap closings on both sides in the magnetic field case.

\begin{figure*}
\begin{center}
\includegraphics[width=1.75\columnwidth, clip]{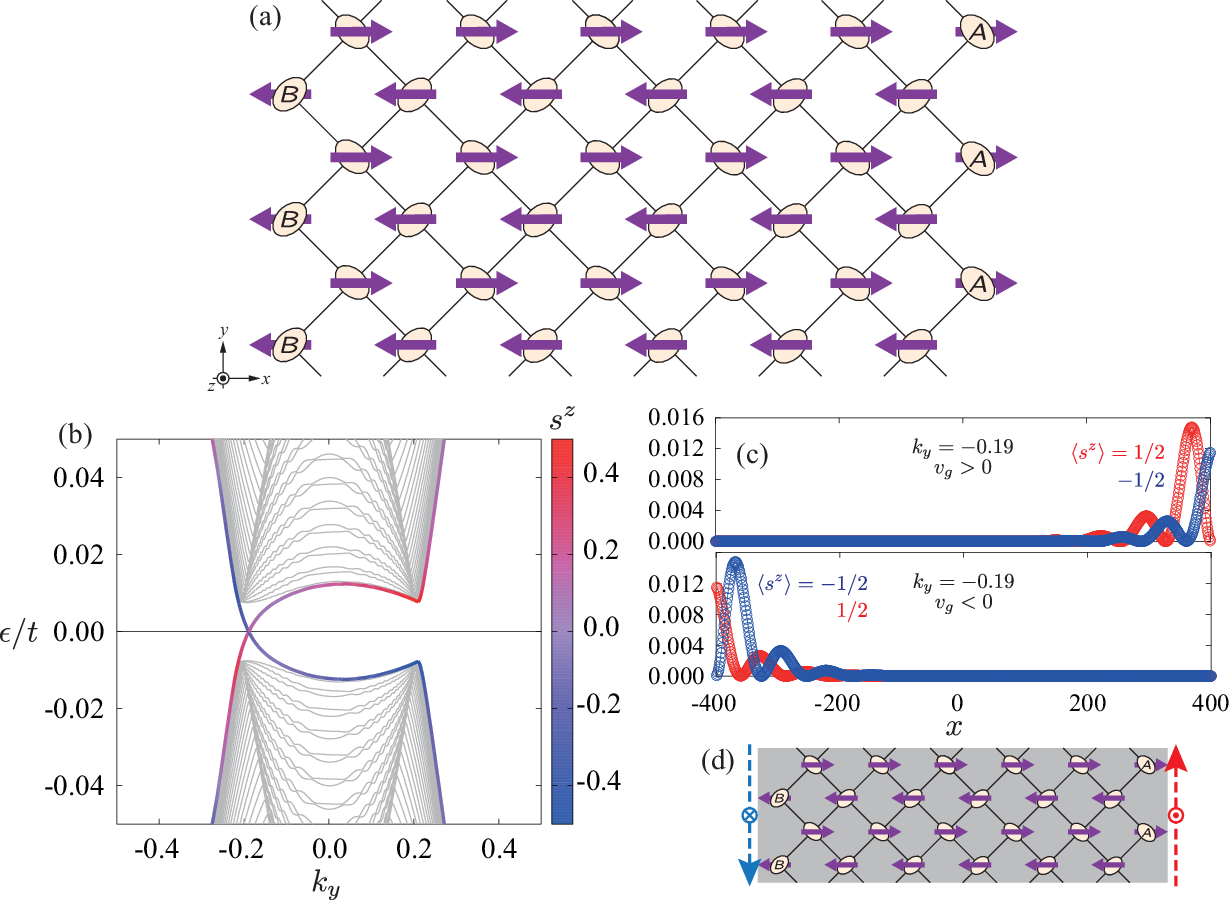}
\end{center}
\caption{
(a) Schematic illustration of the ribbon geometry used in the edge-state calculations for the topological altermagnetic state. 
Open and periodic boundary conditions are imposed along the $x$ and $y$ directions, respectively. 
(b) Energy spectrum as a function of $k_y$ in the topological altermagnetic state induced by the staggered potential. 
The gray curves represent the bulk states, and the colored curves denote the in-gap edge states. 
The color scale indicates the expectation value of the spin moment $\langle s^z \rangle$. 
(c) $s^z$-resolved real-space distributions of the chiral edge modes at $k_y=-0.19$ in panel (b).
The spin-up and spin-down components are shown by red and blue symbols, respectively. 
The upper (lower) panel shows the edge state with positive (negative) group velocity $v_g$. 
(d) Schematic illustration of the edge states shown in panels (b) and (c). 
The dashed arrows along the edges indicate the propagation directions of the edge modes, while the out-of-plane arrows denote the sign of the spin polarization $\langle s^z \rangle$.
}
\label{fig5}
\end{figure*}

\subsection{Edge states}
To further characterize the two topological altermagnetic phases with different quantized Hall conductivities, we next examine their edge states. 
Since the bulk gap closings associated with the topological transitions occur along the $M$-$X$-$M$ line, we employ a ribbon geometry that is open along the $x$ direction and periodic along the $y$ direction, as illustrated in Fig.~\ref{fig5}(a). 
The system consists of $N_x=800$ and $N_y=400$ unit cells along the $x$ and $y$ directions, respectively.
In the following calculations, to fix the antiferromagnetic pattern, we replace the Coulomb interaction term in ${\cal H}_{\rm ALM}$ by an antiferromagnetic molecular field, $- h_{\rm AF}^x \left( \sum_{i \in A} - \sum_{i \in B} \right) s_i^x$. 
The hopping, the SOC, and the external field terms are taken to be the same as those used in the previous self-consistent bulk calculations.

\subsubsection{C=1 state under staggered potential}
Figure~\ref{fig5}(b) shows the energy spectrum of the topological altermagnetic phase with $C=1$ induced by the staggered potential with $\Delta/t = 0.01$ for $h^x_{\rm AF}/t = 0.2$ and $\lambda^A/t = \lambda^R/t = 0.1$. 
The gray curves represent the bulk bands, which show an insulating gap opened by the molecular field $h^x_{\rm AF}$.
The colored curves correspond to the in-gap edge states, which cross the Fermi level only in the $k_y<0$ region; the color scale indicates the expectation value of the spin moment $\langle s^z \rangle$. 
This is consistent with the bulk band structure under periodic boundary conditions shown in Fig.~\ref{fig3}(f), where the gap closing occurs only along the $M$-$X$ line in the $k_y<0$ region.
Two edge-state branches are observed within the bulk gap: one is strongly polarized toward positive $\langle s^z \rangle$, while the other is toward negative $\langle s^z \rangle$. 
In contrast, the spin polarization along the $x$ direction is much smaller, approximately one-fifth of the $s^z$ component, and the $s^y$ component is zero (not shown).

The sign of the spin polarization is correlated with the propagation direction of the edge states. 
The branch with positive (negative) $\langle s^z \rangle$ possesses a positive (negative) group velocity as shown in Fig.~\ref{fig5}(b). 
Figure~\ref{fig5}(c) shows the real-space probability distributions of the edge states near the Fermi level, resolved into branches with positive and negative $\langle s^z \rangle$. 
The mode with positive (negative) group velocity is localized at the right (left) edge of the ribbon.
Thus, the right (left) edge state has the positive(negative)-$s^z$ component, consistent with the spin polarizations of the corresponding Bloch states in the $k_y$ space.

The real-space picture of the edge states is schematically illustrated in Fig.~\ref{fig5}(d). 
Electrons with positive $\langle s^z \rangle$ propagate along the $+y$ direction at the right boundary, while those with negative $\langle s^z \rangle$ propagate along the $-y$ direction at the left boundary. 
These edge states constitute a single chiral edge channel, consistent with the Chern number $C=1$ obtained from the bulk calculations.

\begin{figure*}
\begin{center}
\includegraphics[width=1.75\columnwidth, clip]{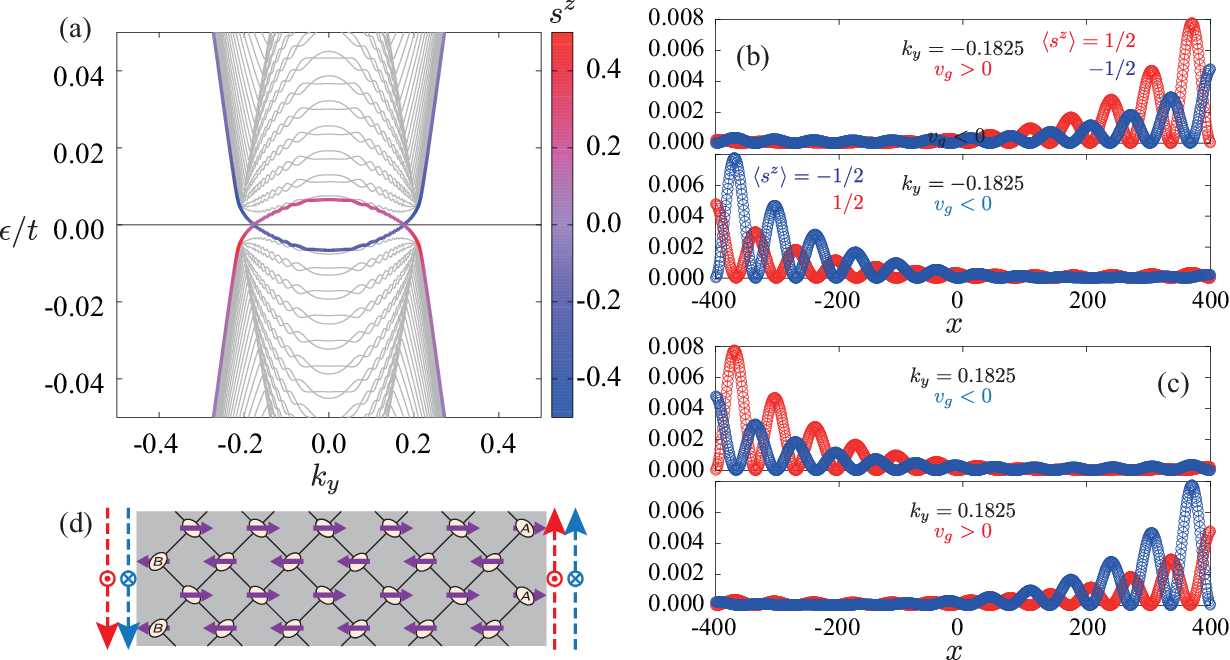}
\end{center}
\caption{
(a) Energy spectrum as a function of $k_y$ in the topological altermagnetic state induced by the magnetic field.
The gray curves represent the bulk states. The color scale for the in-gap edge states indicates the expectation value of the spin moment $\langle s^z \rangle$.
(b), (c) $s^z$-resolved real-space distributions of the chiral edge modes at $k_y=-0.1825$ and $k_y=0.1825$, respectively.
The spin-up and spin-down modes are shown by red and blue symbols.
The upper and lower panels, respectively, show the two edge states crossing the Fermi level, and their signs of the group velocities $v_g$ are indicated in the panels.
(d) Schematic illustration of the edge states.
The dashed arrows along the edges indicate the propagation directions of the edge modes, while the out-of-plane arrows denote the sign of the spin polarization $\langle s^z \rangle$.
}
\label{fig6}
\end{figure*}

\subsubsection{C=2 state under magnetic field}
Next, we examine the edge states in the topological altermagnetic metastable state with $C=2$, realized under the uniform magnetic field $h$ applied along the $z$ direction. 
Figure~\ref{fig6}(a) shows the energy spectrum obtained from the ribbon geometry calculation at $h/t=0.01$, $h^x_{\rm AF}/t = 0.2$, and $\lambda^A/t = \lambda^R/t = 0.1$. 
As in Fig.~\ref{fig5}(b), the bulk bands are shown by gray curves and the edge states are highlighted by the color scale representing the expectation value of the spin moment $\langle s^z \rangle$.

In contrast to the staggered-potential-induced $C=1$ state, where gapless edge states appear only in the $k_y<0$ region, the present $C=2$ state exhibits gapless edge modes in both the $k_y<0$ and $k_y>0$ regions. 
This behavior is consistent with the bulk band structures under periodic boundary conditions, where the topological transition is accompanied by simultaneous gap closings on the both sides of the $X$ point, as shown in Fig.~\ref{fig4}(d). 
As a result, four edge-state branches appear inside the bulk gap, with two branches in each of the $k_y<0$ and $k_y>0$ sectors. 
Similar to the $C=1$ state, these edge modes are strongly polarized along the $z$ direction.

The relation between the spin polarization and the propagation direction depends on the sign of $k_y$. 
In the $k_y<0$ region, the edge states exhibit the same behavior as those in the $C=1$ state: the modes with positive (negative) $\langle s^z \rangle$ possess positive (negative) group velocities. 
In contrast, this correspondence is reversed in the $k_y>0$ region, where the modes with positive (negative) $\langle s^z \rangle$ have negative (positive) group velocities.

Figures~\ref{fig6}(b) and \ref{fig6}(c) show the $s^z$-resolved real-space distributions of the edge modes in the $k_y<0$ and $k_y>0$ regions, respectively. 
For $k_y<0$, the mode with positive $\langle s^z \rangle$ and positive group velocity is localized at the right edge, whereas the mode with negative $\langle s^z \rangle$ and negative group velocity is localized at the left edge, analogous to the $C=1$ state. 
For $k_y>0$, however, the localization pattern is reversed: the right edge hosts the mode with negative $\langle s^z \rangle$ and positive group velocity, while the left edge hosts the mode with positive $\langle s^z \rangle$ and negative group velocity. 
We note that the weaker localization of the edge modes compared with the $C=1$ state reflects the smaller bulk gap in the present magnetic-field-induced state.

The resulting real-space picture is summarized schematically in Fig.~\ref{fig6}(d). 
Unlike the $C=1$ state, where only a single spin-polarized edge mode appears on each boundary, the present phase hosts two edge modes with opposite spin polarizations but identical propagation direction on each edge. 
The appearance of two chiral edge modes at each boundary is consistent with the Chern number $C=2$ obtained from the bulk calculations.

\section{Discussion}
Now let us discuss the experimental relevance of our results.
The present study suggests two promising routes toward the experimental realization of QAHE in altermagnets.
A common ingredient in both routes is the Rashba-type SOC, which induces the band inversion underlying the topological phase transition.
Therefore, altermagnetic thin films, interfaces, and heterostructures with broken inversion symmetry provide natural platforms for realizing the proposed mechanisms.
The required strengh of Rashba coupling is primarily determined by the magnitude of the antiferromagnetic insulating gap and is typically of the order of $\lambda^{\rm R}/t \sim 0.1$.
Such a coupling strength is expected to be achievable at transition-metal-oxide interfaces involving $3d$, $4d$, and $5d$ transition-metal ions.
In molecular materials, such a coupling strength is generally difficult to achieve in purely organic systems.
However, organic--metal complexes with hybridized $\pi$ and $d$ orbitals containing heavy transition-metal ions may provide another promising platform.
For example, recently synthesized gold dithiolene molecular conductors based on diethoxybenzene-substituted ligands exhibit canted antiferromagnetic order together with $\pi$--$d$ hybridized electronic states~\cite{Yokomori2026}.
This suggests that molecular $\pi$--$d$ systems containing heavy $5d$ transition-metal ions could be attractive candidates for realizing the present mechanism.

For the first route, the staggered potential can be introduced naturally in altermagnets through lattice distortions.
While introducing such a staggered potential is generally difficult in conventional antiferromagnets with a single magnetic site in the crystallographic unit cell, it can be readily realized in altermagnets by lattice distortions that break the symmetry relating the crystallographically equivalent spin-up and spin-down sublattices.
The simplest realization is provided by uniaxial compression or tension along the $[110]$ or $[1\bar10]$ directions.
Notably, this lattice distortion is identical to that responsible for the piezomagnetic effect in altermagnets~\cite{Aoyama2024, Yershov2024, Naka2025}. 
In addition, since the resulting piezomagnetic state is symmetry-equivalent to a so-called compensated ferrimagnet~\cite{Naka2025}, compensated ferrimagnetic thin films may also provide a promising platform for realizing the quantized Hall state without applying external strain.

The second route exhibits a markedly different character from the first one, in that the quantized anomalous Hall state appears not in the ground state but in a metastable state. 
In this sense, the present mechanism realizes a hidden topological phase that is inaccessible under equilibrium conditions. 
Although metastable states are generally challenging to prepare and control experimentally, the metastable state discussed here naturally appears during the magnetic hysteresis process associated with the reversal of the weak ferromagnetic moment. 
Such hysteretic metastable states are ubiquitous in magnetic materials and can be accessed by conventional magnetization processes, even in molecular systems with relatively small magnetic anisotropy. 
This makes the present mechanism a realistic route toward realizing metastable quantum anomalous Hall states.

Finally, the present study highlights that the characteristic symmetry of altermagnets provides unique opportunities for realizing topological quantum states through experimentally controllable external perturbations. 
We expect that the interplay among altermagnetic order, SOC, and band topology will offer a fertile platform for exploring new quantum transport phenomena in magnetic materials.

\begin{acknowledgments}
This work is supported by Grant-in-Aid for Scientific Research, No. 23K25826, 23K03333, 25H00838, 23K13056, 25H01247, and 26H00636.
\end{acknowledgments}



\end{document}